\documentclass[prb,twocolumn,floats,citeautoscript]{revtex4}

\usepackage{graphicx}
\usepackage{amsmath}
\usepackage{amssymb}
\usepackage{times}
\usepackage{colordvi}

\graphicspath{{.}{./EPS/}}

\newcommand{\ket}[1]{\left | \, #1 \right \rangle}

\begin{document}

\title{Tunable entanglement generation for mobile--electron spin qubits}

\author{A.~I. Signal}
\affiliation{Institute of Fundamental Sciences, Massey University,
Private Bag 11~222, Palmerston North, New Zealand}

\author{U. Z\"ulicke}
\email{u.zuelicke@massey.ac.nz}
\affiliation{Institute of Fundamental Sciences and MacDiarmid Institute for
Advanced Materials and Nanotechnology, Massey University,
Private Bag 11~222, Palmerston North, New Zealand}

\date{\today}

\begin{abstract}

Recent studies have shown that linear electron optics can be used to
generate entangled two--particle states from nonentangled ones if additional
measurements of charge or parity are performed. We have investigated 
such nondeterministic entanglement production in electronic versions of the
Mach--Zehnder interferometer, where spin--dependent interference occurs
due to the presence of electric--field tuneable Rashba spin splitting.
Adjustment of the spin--precession length turns out to switch the
entangler on and off, as well as control the detailed form of entangled
output states.

\end{abstract}

\pacs{73.43.Cd, 73.43.Jn}

\maketitle

Entanglement is seen as an important resource for quantum information
processing (QIP), enabling exponential speed-up of classically hard
computations, secure cryptography, and ultradense coding of
information~\cite{qcqip}. Generating entanglement in the electronic spin
degree of freedom is therefore essential for many solid--state--based QIP
schemes~\cite{lossbook}. Several proposals rely on two--particle correlation
effects induced, e.g., by superconducting leads~\cite{supentang}, Coulomb
blockade in quantum dots~\cite{dotentang}, or quasiparticle
collisions~\cite{sara:prl:04}. It is, however, possible to obtain
spin--entangled two--electron states in the absence of interactions using
scattering interference at beam splitters~\cite{bos:prl:02,qpcentang,
been:prl:04}. This opens up a promising avenue for realizing
QIP based on linear fermion optics~\cite{been:prl:04} as a solid--state
analog of efficient quantum computation with photons~\cite{milb:nat:01}.

Our present study combines recent insight into linear--optics QIP with the
possibilities of spin--dependent electron interference~\cite{nitta:apl:99,
uz:apl:04}. We find that maximally entangled mobile electron pairs can be
extracted at the output of an electronic Mach--Zehnder (MZ) interferometer,
realized in an asymmetric semiconductor heterostructure, using a projective
charge measurement. The efficiency of entanglement generation depends
on the spin--precession length $L_{\text{so}}$ for electrons in the 
interferometer arms and is therefore tuneable by external gate voltages.
The same is true for the detailed form of entangled two--qubit states that are
generated. Hence it is possible to accomplish electric--field control of
entanglement production and entangled--state output in the device
considered here.

The basic setup and transport  properties of a spin--dependent electronic
MZ interferometer were studied
theoretically
in Ref.~~\onlinecite{uz:apl:04}.
(See Ref.~~\onlinecite{electronMZ} for a recent experimental
realization of a MZ interferometer {\em without\/} spin dependence.)
It is a four--terminal (two--input, two--output) device where electron--wave
interference depends on spin due to the presence of Rashba spin
splitting~\cite{byra:jpc:84}. Linear conductances were calculated based on a
spin--resolved single--electron scattering matrix for the entire interferometer.
For our present study, we investigated \textit{two--electron\/} interference in
this system, which is a nontrivial generalization because the Pauli principle
affects the outcome of multi--particle scattering events~\cite{loud:pra:98}.
Two--particle interference at beam splitters can be used for detecting
entanglement~\cite{burkloss:prb-rc:00}, and spin precession from Rashba
spin splitting offers a way to manipulate entangled states~\cite{egues:prl:02}.
We proceed now to briefly sketch our theoretical description.

The two--particle Hilbert space for electrons at any four--terminal device is
six--dimensional. We find it useful to use a magic basis of Bell
states~\cite{eck:ann:02}
\begin{subequations}
\begin{eqnarray}
\ket{\chi_1}&=&\frac{1}{\sqrt{2}}\left(c^\dagger_{\mathrm{a}+}
c^\dagger_{\mathrm{a}-} + c^\dagger_{\mathrm{b}+} 
c^\dagger_{\mathrm{b}-} \right) \ket{0} \quad , \\ 
\ket{\chi_2}&=&\frac{1}{\sqrt{2}}\left(c^\dagger_{\mathrm{a}+} 
c^\dagger_{\mathrm{b}+} - c^\dagger_{\mathrm{a}-} 
c^\dagger_{\mathrm{b}-} \right)  \ket{0} \quad , \\
\ket{\chi_3}&=&\frac{1}{\sqrt{2}}\left(c^\dagger_{\mathrm{a}+} 
c^\dagger_{\mathrm{b}-} + c^\dagger_{\mathrm{a}-} 
c^\dagger_{\mathrm{b}+} \right)  \ket{0} \quad , \\
\ket{\chi_4}&=&\frac{i}{\sqrt{2}}\left(c^\dagger_{\mathrm{a}+} 
c^\dagger_{\mathrm{a}-} - c^\dagger_{\mathrm{b}+}
c^\dagger_{\mathrm{b}-} \right)  \ket{0} \quad , \\
\ket{\chi_5}&=&\frac{i}{\sqrt{2}}\left(c^\dagger_{\mathrm{a}+} 
c^\dagger_{\mathrm{b}+} + c^\dagger_{\mathrm{a}-} 
c^\dagger_{\mathrm{b}-} \right)  \ket{0} \quad , \\
\ket{\chi_6}&=&\frac{i}{\sqrt{2}}\left(c^\dagger_{\mathrm{a}+} 
c^\dagger_{\mathrm{b}-} - c^\dagger_{\mathrm{a}-} 
c^\dagger_{\mathrm{b}+} \right)  \ket{0} \quad ,
\end{eqnarray}
\end{subequations}
where the operator $c^\dagger_{\alpha\sigma}$ creates an electron at the
Fermi energy in the interferometer arm $\alpha$ and with Rashba
spin--subband quantum number $\sigma$. Any two--electron state $\ket
{\text{i}} $ ($\ket{\text{o}}$) at the input (output) can be expressed as a
linear combination of these basis states. It is straightforward to derive the
$6\times  6$ two--particle scattering matrices ${\mathcal S}_{\text{2el}}$ that 
relate two--particle input and output states for each linear--optics element
(such as beam splitters, mirrors, and the spin--dependent phases acquired
when the electrons propagate through interferometer arms). As examples,
we give the one for an
ideal symmetric
beam splitter,
\begin{equation}
{\mathcal S}_{\text{2el}}^{\text{(bs)}}=\left(
\begin{array}{llllll}
 0 & \frac{1}{\sqrt{2}} & 0 & 0 & 0 & \frac{1}{\sqrt{2}}
   \\
 \frac{1}{\sqrt{2}} & -\frac{1}{2} & 0 & 0 & 0 &
   \frac{1}{2} \\
 0 & 0 & -1 & 0 & 0 & 0 \\
 0 & 0 & 0 & -1 & 0 & 0 \\
 0 & 0 & 0 & 0 & -1 & 0 \\
 -\frac{1}{\sqrt{2}} & -\frac{1}{2} & 0 & 0 & 0 &
   \frac{1}{2}
\end{array}
\right) \quad ,
\end{equation}
and the one describing propagation through inner interferometer arms
(horizontal and vertical device dimensions given by $x$ and $y$,
respectively) where the spin--precession length enters:
\begin{widetext}
\begin{equation}
{\mathcal S}_{\text{2el}}^{\text{(Ra)}}=\left(
\begin{array}{llllll}
 1 & 0 & 0 & 0 & 0 & 0 \\
 0 & -\cos \left(\frac{\pi  (x+y)}{L_{\text{so}}}\right) & 0 & 0 &
   \sin \left(\frac{\pi  (x+y)}{L_{\text{so}}}\right) & 0 \\
 0 & 0 & -\cos \left(\frac{\pi  (x-y)}{L_{\text{so}}}\right) & 0 & 0 &
   \sin \left(\frac{\pi  (x-y)}{L_{\text{so}}}\right) \\
 0 & 0 & 0 & -1 & 0 & 0 \\
 0 & -\sin \left(\frac{\pi  (x+y)}{L_{\text{so}}}\right) & 0 & 0 &
   -\cos \left(\frac{\pi  (x+y)}{L_{\text{so}}}\right) & 0 \\
 0 & 0 & \sin \left(\frac{\pi  (x-y)}{L_{\text{so}}}\right) & 0 & 0 &
   \cos \left(\frac{\pi  (x-y)}{L_{\text{so}}}\right)
\end{array}
\right) \quad .
\end{equation}
\end{widetext}
Suitable combination of the respective matrices for beam splitters, mirrors,
etc.\  yields the scattering matrix
${\mathcal S}_{\text{2el}}^{\text{(MZI)}}$ for the entire MZ interferometer,
which depends on the device dimensions measured in units of the
spin--precession length. Its analytical form is unilluminating, hence we omit it
here. With this result, we can discuss two--electron scattering interference at
spin--dependent MZ interferometers in full generality.

\begin{figure}[b]
\includegraphics[width=3.in]{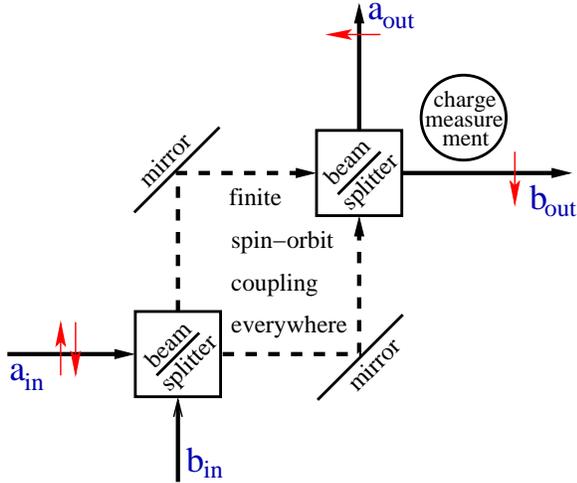}
\caption{(Color online) Entanglement generation at a spin--dependent
Mach--Zehnder interferometer. A two--particle product state is incident in
channel a, as indicated. The output state generated by the interferometer is
projected, by a charge measurement, onto its component with single electron occupancy in each output arm. The resulting two--electron state turns out to
be maximally entangled in the electrons' quantum number for spin projection
perpendicular to their respective interferometer arm, i.e., Rashba
spin-subband index.
\label{fig1}}
\end{figure}
\begin{figure}[b]
\includegraphics[width=3.in]{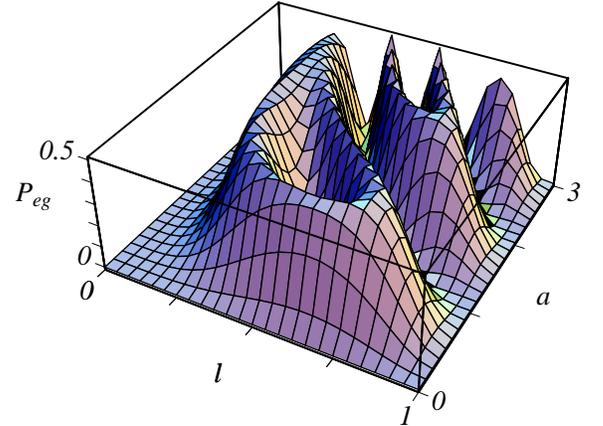}
\caption{(Color online) Efficiency 
$P_{\text{eg}}$
of entanglement generation by the procedure illustrated in Fig.~\ref{fig1},
calculated for rectangular MZ interferometers of width $l L_{\text{so}}$ and
aspect ratio $a$.\label{fig2}}
\end{figure}
We focus on a specific situation where the input is a two--particle product
state with double--occupancy in one interferometer arm. See 
Fig.~\ref{fig1} for an illustration. Without loss of generality, we choose the
state $\ket{\text{i}_{\text{dbl}}} = c^\dagger_{\mathrm{a}+} 
c^\dagger_{\mathrm{a}-}\ket{0}$. In general, the
output state generated from it by the MZ interferometer will be a quantum
superposition of states having single electron occupancy in each output arm
with others that have a finite amplitude for double occupancy in one of them.
Lets assume we have the  means to measure the electron number (i.e.,
charge) in one of the output channels (e.g., in arm b). Performing this
measurement each time we send in two electrons with opposite spin at the
input channel a, we will obtain the values 0, 1, or 2 with certain frequencies.
The charge measurement leaves the spin and orbital wave functions of
output states unaffected, but it allows us to filter out, e.g., such states where
exactly one electron is scattered into each of the output arms. We have
calculated the concurrence~\cite{woot:prl:98} (i.e., a measure of
entanglement) for the single--occupancy output states that are generated
from the double--occupancy input state $\ket{\text{i}_{\text{dbl}}}$ given
above and found that they are \textit{always maximally entangled\/},
irrespective of interferometer geometry and size. Hence, preparing the input
state  $c^\dagger_{\mathrm{a}+} c^\dagger_{\mathrm{a}-} \ket{0}$, which
can be achieved by simply raising the voltage in the electron reservoir
connected to input a by a certain amount, and choosing states at the
interferometer output for which exactly one electron lives in each output
arm, we obtain two--electron states that are maximally entangled in the
Rashba spin quantum number $\sigma$. A schematic illustration of this
procedure is shown in Fig.~\ref{fig1}. Our analysis shows that the same
entangled output state is generated when the two--electron input is incident
in interferometer arm b instead. To perform the crucial charge measurement,
mesoscopic electrometers such as single--electron transistors or quantum
point contacts could be employed~\cite{been:prl:04}.

The efficiency $P_{\text{eg}}$ for nondeterministic entanglement generation as described above is obviously given by the probability of single--electron
occupancy per output arm for the state ${\mathcal 
S}_{\text{2el}}^{\text{(MZI)}} \ket{\text{i}_{\text{dbl}}}$. This quantity depends
on the interferometer size and geometry but turns out to be the same for a
double--occupancy state incident in channel b. In the following, we consider
a rectangularly shaped MZ interferometer of width $w$ and height $h$,
using the dimensionless parameters $l=w/L_{\text{so}}$ and $a=h/w$ for its characterization. We find
\begin{eqnarray}\label{engenprob}
P_{\text{eg}}&=& \frac{1}{2}-\frac{1}{8} \big\{\sin ^2([a-1] l \pi )+\sin
   ^2([a+1] l \pi ) \nonumber \\  && \hspace{1.5cm} +\cos (2 l \pi )+
   \cos (2 a l \pi )\big\}^2 \quad .
\end{eqnarray}
Figure~\ref{fig2} illustrates the existence of maxima in $P_{\text{eg}}$ as a
function of $l$ for given finite $a$. As can be seen from 
Eq.~(\ref{engenprob}), the largest possible value of $P_{\text{eg}}$ is 50\%,
which is the same as the efficiency of entanglement generation at a single
beam splitter~\cite{bos:prl:02}. Unlike the beam splitter, however, the MZ
interferometer can be tuned by an external gate voltage (which controls the
parameter $l$ by adjusting\cite{lsoadjust} $L_{\text{so}}$) between
maximum and zero entanglement generation, realizing a switchable
entangler. Further investigation may show possibilities for increasing the
entanglement--generation efficiency by cascading several MZ
interferometers in series~\cite{bos:prl:02}.
Any backscattering incurred from the charge measurement is detrimental 
for the output of entangled electron pairs and will have to be minimized, e.g.,
by using a highly sensitive~\cite{pepp:prl:93} and efficient~\cite{butt:prl:02}
quantum--point--contact electrometer.

\begin{figure}[t]
\includegraphics[width=3.in]{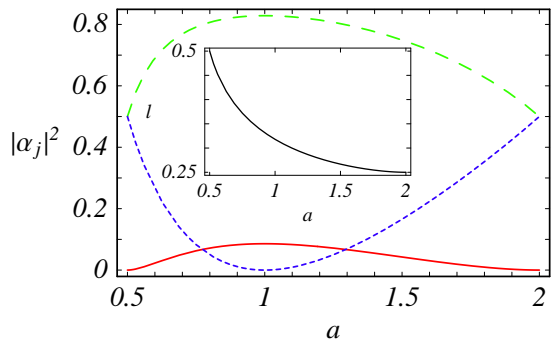}
\caption{(Color online) Admixture of Bell basis states in the output state
generated for interferometer configurations with maximal efficiency of
entanglement generation. The
solid
curve shows the overlap with $\chi_2$, which is identical to that with the
%
state $\chi_6$. The overlap with states $\chi_3$ and $\chi_5$ is given by
the
short--dashed and long--dashed
curves, respectively. Inset: Smallest value of dimensionless interferometer
width $l$ for given aspect ratio $a$ at which the maximum
entanglement--generation efficiency of 50\% is realized.
\label{fig3}}
\end{figure}
We proceed to examine in greater detail the entangled state generated by
the above procedure. Due to the projection, it lives in the subspace spanned
by the Bell states $\chi_2$, $\chi_3$, $\chi_5$, and $\chi_6$. We find for
the respective amplitudes
\begin{subequations}
\begin{eqnarray}
\alpha_2&=&-\alpha_6=\sin(2 l \pi)\sin(2 a l \pi)/(4\sqrt{P_{\text{eg}}})\quad ,
\\ \alpha_3&=&\sin(l\pi)\sin([a-1]l\pi)\sin(a l \pi)/ \sqrt{P_{\text{eg}}} \quad , \\
\alpha_5&=&\sin(l\pi)\sin([a+1]l\pi)\sin(a l \pi)/ \sqrt{P_{\text{eg}}} \quad .
\end{eqnarray}
\end{subequations}
As external gate voltages can adjust the value of $l$, it is possible to
control the detailed form of entangled output states generated by a given
interferometer setup. Situations where the efficiency of entanglement
production reaches its maximal value of 50\% are of greatest practical
interest, hence we investigate those further. As Fig.~\ref{fig2} indicates,
there may exists more than one value of dimensionless interferometer width 
$l$ for which $P_{\text{eg}}$ is maximized at a given aspect ratio $a$. To be 
specific, we consider always the smallest of such $l$ values in the following,
as plotted in the inset of Fig.~\ref{fig3}. The overlap of maximum--efficiency
output states with Bell basis states is shown in Fig.~\ref{fig3}. Bell states
that are symmetric under spin reversal generally provide the largest
contribution.

In conclusion, we presented a procedure for generating entangled electron
pairs using spin--dependent interferometry and a projective charge
measurement. Electric--field control of the spin precession length enables
manipulation of the entangled ouput state and realization of a switchable
entangler.

We benefitted from useful discussions with M.~Governale, J.~K\"onig, and
Y.~Tokura. UZ gratefully acknowledges support from the Marsden Fund of
the Royal Society of New Zealand.


\end{document}